\documentclass[epj,final]{svjour}
\usepackage[dvipsnames]{xcolor}
\usepackage{graphicx}
\usepackage{float}
\usepackage{textcomp}
\usepackage{latexsym}
\usepackage{url}
\usepackage{amsfonts}
\usepackage{amsmath, amssymb}
\RequirePackage{graphicx}
\usepackage{caption,subcaption}
\usepackage{indentfirst}
\usepackage{svg}

\begin{document}
	\title{S2-star dynamics probing the Galaxy core cluster}
        \titlerunning{Neural Network Analysis of S2-Star Dynamics}
	\author{N.Galikyan\inst{1,2}, Sh.Khlghatyan\inst{2}, A.A.Kocharyan\inst{3}, V.G.Gurzadyan\inst{2,4}
	}                     
	%
	%
	\institute{National Research Nuclear University MEPhI, Moscow, Russia \and Center for Cosmology and Astrophysics, Alikhanian National Laboratory and Yerevan State University, Yerevan, Armenia  \and School of Physics and Astronomy, Monash University, Clayton, Australia \and SIA, Sapienza Universita di Roma, Rome, Italy}
	\date{Received: date / Revised version: date}
	%

	\abstract{The star cluster surrounding the supermassive black hole in the center of Milky Way is probed using the data on the S2 star. The value of precession found at the physics-informed neural networks (PINN) analysis of the S2 data is used to consider the role of the scattering of S2 star on stars of the cluster, described by a random force given by the Holtsmark distribution. The critical value for the star density of the core cluster for which the observed precession value by PINN lies inside 70\% confidence interval (between $15\%$ and $85\%$ quantiles) around the median of precession due to scattering, is obtained as $n_{\text{crit}}\approx 8.3\times 10^{6}\, pc^{-3}$, that is at higher star densities the perturbation of the orbit of S2 would exceed the observed one.  
}

%
\maketitle

\section{Introduction}

Galactic centers have long been among the major goals of intense observational and theoretical investigations. The presence of a central massive black hole surrounded by a dense star cluster, the accretion onto the black hole, the stellar tidal disruption events and the diversity of linked physical effects are traced at observational dedicated surveys. The revealing of the black hole shadows in the centers of the galaxy M87 and of the Milky Way had enabled to probe the properties of the metric in the vicinity of the central black hole \cite{Event1,Event2}.   

The monitoring of S-stars motion around the supermassive black hole (SMBH) SgrA* in the center of the Galaxy represents a unique window to test the General Relativity (GR), i.e. the motion of S2 star enabled to detect the precession consistent with the predictions of GR for the Schwarzschild metric \cite{GRAVITY}. The data on the S-stars have been used to constrain modified theories of gravity, the dark matter models as scalar and vector clouds, the extended mass component in the gravitational potential, etc, (see  \cite{Cap,Fermionic,Borka,Z1,Z2,Yukawa,EinMax,GRAVITYscalarclouds,GRAVITYvector}), thus complementing the other tests of gravity theories, e.g. \cite{Nuc,Z3,Ciu1,Ciu2,Ciu3,Ciu4,SKG}.

The observational data on the S-stars \cite{Gil} have been analyzed in \cite{PINNSstars,PINNextended} by means of the physics-informed neural networks (PINN) \cite{PINN1,PINN2}. In \cite{PINNSstars} the weak-field modified GR \cite{G,G1,GS1} involving the cosmological constant  has been constrained. The interest to that modified gravity version was due to its fitting the data on the dynamics of groups and clusters of galaxies \cite{G1,GS1,GS2}, providing an explanation to the Hubble tension as a  result of two flows, local and global ones \cite{GS3,GS4}, to the properties of the filaments in the local Universe \cite{GFC1,GFC2,GFC3}. In \cite{PINNextended} the PINN was used to constrain the extended dark matter contribution to the S2 precession and the conclusion on the absence of a signature of extended mass up to 0.01\% of the central mass inside the apocenter of S2 was drawn.

In the present study we use the S2 data and the PINN to probe the parameters of the star cluster surrounding the central black hole. The estimates on the star density in the Galactic core e.g. at a distance of $0.01\, pc$ from center were around $n \simeq 2.6 \pm 0.3\,\times 10^7\, pc^{-3}$ \cite{Sh} , hence, it is of particular interest to probe the central stellar cluster role using the S2 data. We used the Holtsmark distribution approach $H(\beta)$  on the statistics of the gravitational force acting on a star within uniformly distributed point masses \cite{ChandraStoch,ChNeu}
\begin{equation}
H(\beta)=\frac{2}{\pi\beta}\int_0^{\infty}x \sin x e^{-(\frac{x}{\beta})^{3/2}} dx,
\end{equation}
where $H(\beta)d\beta$ is the probability for the relative force  $\beta=\varepsilon/\varepsilon_{\text{mean}}$; $\varepsilon_{\text{mean}}$ within the interval $d\beta$ is given below, Eq.(\ref{eq:Holtsmark}). This approach assumes the random character of the scatter of the monitored star S2 on a stars of the star cluster. The details of the scatter of the gravitating bodies are discussed in \cite{ChandraStoch}.
Thus, analysing the possible influence (scatter) of galactic core stars on the precession of S2-star, we reveal the critical i.e. the maximal star concentration $n_{\text{crit}}\approx 8.3\times 10^{6}\, pc^{-3}$ which could result in the observed precession, while at higher core's star concentration the orbit of S2 would have been disturbed more than allowed by the observations.  

\section{PINN}

\begin{figure}[H]
    \centering
    \includegraphics[width=0.8\linewidth]{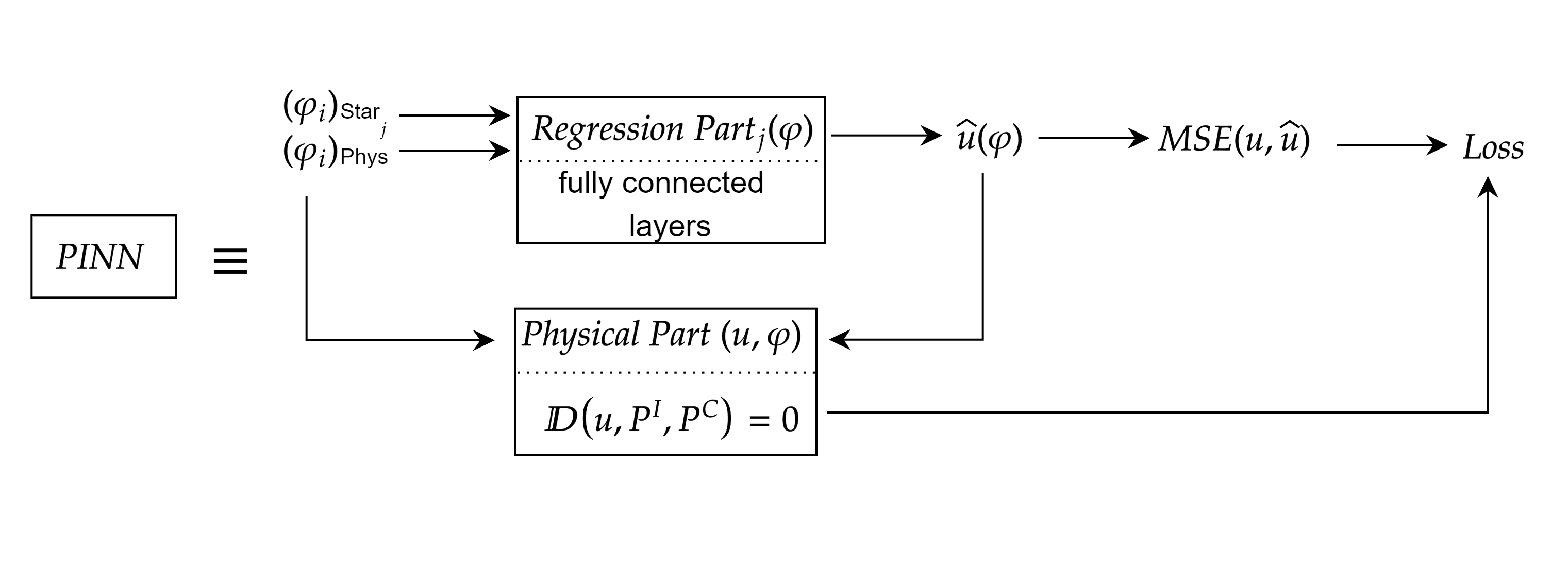}
    \caption{PINN basic scheme}
    \label{fig:scheme_0}
\end{figure}

\begin{figure}[H]
    \centering
    \includegraphics[width=0.8\linewidth]{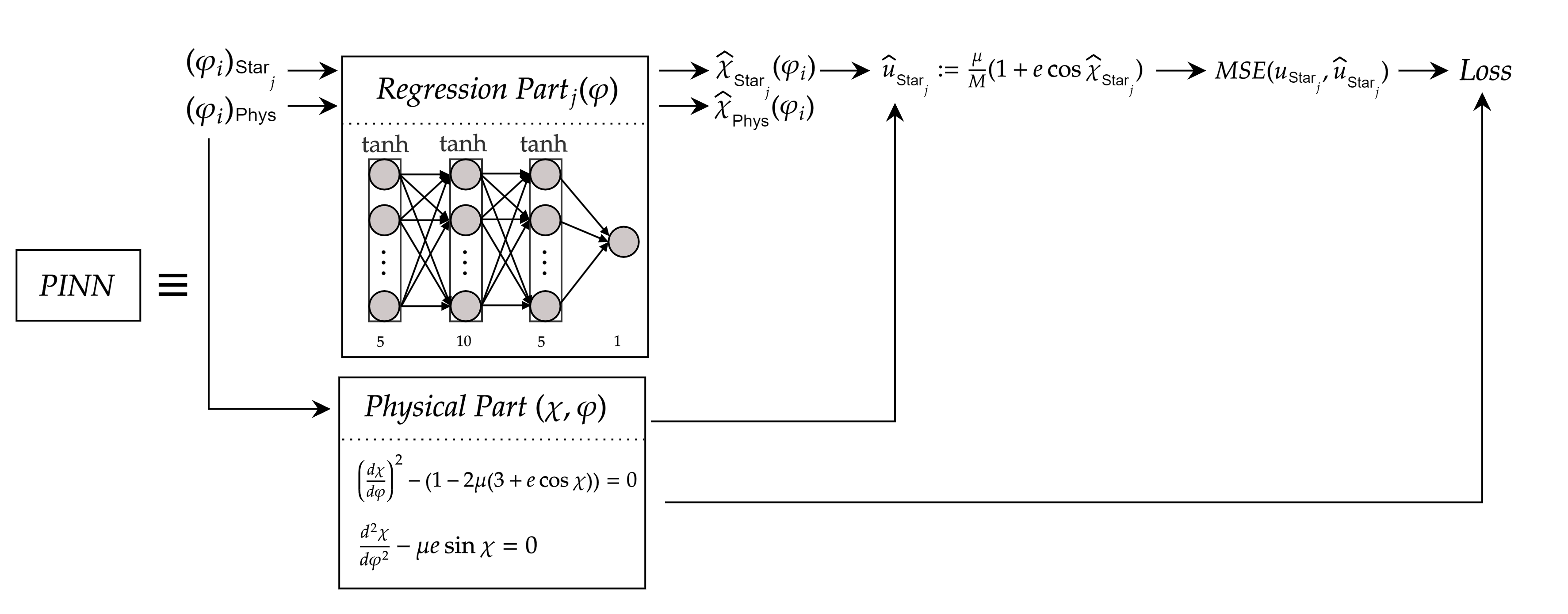}
    \caption{Darwin PINN scheme}
    \label{fig:scheme_1}
\end{figure}

In this section we briefly describe the PINN and highlight the previous results \cite{PINNSstars} needed for this study.
The PINN architecture consists of two parts  as shown in Fig.(\ref{fig:scheme_0}):
\begin{itemize}
    \item A neural network with dense layers in which the input are the polar angles $\varphi$ and output is the inverse radius $u:=\tfrac{1}{r}$. The output is compared with the observed data and minimizes the mean squared error between them.
    \item The physical part includes differential equations corresponding to our physical model and provides an additional contribution to the loss function. The physical process is described by~Eq.(\ref{eq:diff_eq})
\begin{equation} \label{eq:diff_eq}
    F(x, y, y'_x, y_x'', \ldots, y_x^{(n)}) = 0,
\end{equation}
which may have train parameters. Then the physical loss is given by the following loss function Eq.(\ref{eq:phys_loss})
\begin{equation} \label{eq:phys_loss}
    L_{phys}(f(x), x) = F^2\left(x, f(x), f'(x), f''(x), \ldots, f^{(n)}(x)\right).
\end{equation}
\end{itemize}
The physical process can be described by different gravity models. In this paper we use the results of the training based on GR and Darwin's equations of motion \cite{Chandra}
\begin{align}
\frac{d^2\chi}{d\varphi^2}&=\mu e \sin\chi, \label{eq:Darwin_1}\\
\left(\frac{d\chi}{d\varphi}\right)^2&=1-2\mu(3+e\cos\chi), \label{eq:Darwin_2}\\
u =\frac{\mu}{M_{\bullet}}&(1+e\cos\chi),\,\, \mu:=\frac{M_{\bullet}}{p}, \label{eq:Darwin_3}
\end{align}
where $u:=\frac{1}{r}$, $M_{\bullet}$ is the SMBH mass, $e$ and $p$ are respectively the eccentricity of S2 orbit and the focal parameter of the elliptical orbit, and $\chi$ is a variable called a relativistic anomaly \cite{Chandra}. The $e,p$, and $M_{\bullet}$ are the trainable parameters. Using this scheme Fig.(\ref{fig:scheme_1}) for training of the S2 data we found the following values for precession, orbital parameters and SMBH mass (the value of mass was fixed after some point during the training) \cite{PINNSstars}:

\begin{equation}\label{eq:PINN_res}
    \delta\varphi_{\text{Reg}}=11.84 ' \pm 0.03 ', \quad \hat{e}=0.88512 \pm 0.00001 ,\quad \hat{p}=219.2 \pm 0.2\, au,\quad \hat{M}_{\bullet} = 0.04\, au\equiv 4.05\times 10^{6} M_{\odot}.
\end{equation}

\section{Scattering} \label{sec:Scattering}

To find the influence of a cluster of stars in the Galactic center on the motion of S2 star, we first investigate small angle scattering of S2 on the cluster stars. 

Consider the Keplerian motion of the S2 star with parameters of orbit $p$ and $e$,
\begin{equation}
    r = \frac{p}{1+e\cos\varphi},
\end{equation}
and a star with mass $M=M_{\odot}$ at a point $(x_0, y_0, z_0)$ that acts as a scattering center for S2 star. In that case, we assume that a single act of scattering happens at the point $(x_*, y_*, 0)$, where $(x_*, y_*)$ are the coordinates of the intersection of S2 orbit and a line from $(0, 0, 0)$ to $(x_0, y_0)$. The scattering happens with impact parameter $b$
\begin{equation}
    b=\sqrt{z_0^2+(x_0-x_*)^2 + (y_0-y_*)^2},
\end{equation} 
and momentum $\Delta\vec{\rho}$ is transferred to S2, where $\Delta\vec{\rho}$ points from $(x_*, y_*, 0)$ to $(x_0, y_0, z_0)$. Due to the small angle scattering $\Delta\rho$ can be estimated as
\begin{equation}\label{eq:landau}
    \Delta \rho = - \frac{b}{\rho(\varphi_*)}\int_{-\infty}^{\infty}\frac{GM}{(x^2 + b^2)^{3/2}} \,dx = \frac{2GM}{\rho(\varphi_*) b}.
\end{equation}
The energy is increased by $\Delta E = \vec{\rho}\cdot\Delta\vec{\rho}$ and the angular momentum is increased by $\Delta L = \Delta\rho_{\text{ang}} r_{*}$, where $\Delta\rho_{\text{ang}}$ is the non-radial component of transferred momentum. Substituting the momentum $\rho(\varphi_*)$ at point $(r_*, \varphi_*)$, we get 
\begin{equation}\label{eq:delta_p1}
    \Delta \rho = \frac{-2p GM}{bL\sqrt{1+e^2+2e\cos\varphi_*}}.
\end{equation}
After the scattering, we recalculate the values of the focal parameter $\tilde{p}$ and eccentricity $\tilde{e}$ and get an updated law of motion
\[
r = \frac{\tilde{p}}{1 + \tilde{e}\cos(\varphi + \delta\varphi_s)}.
\]
Then we find the value of the additional precession due to scattering $\delta\varphi_s$ by equating the radius of the orbit at point of scattering, i.e.
\begin{equation}\label{eq:delta_phi}
    \frac{p}{1+e\cos\varphi_{*}} = \frac{\tilde{p}}{1 + \tilde{e}\cos(\varphi_{*} + \delta\varphi_s)} . 
\end{equation}

\begin{figure}[H]
    \centering
    \begin{tabular}{cc}
        \begin{subfigure}[b]{0.45\textwidth}
            \centering
            \includegraphics[width=1\textwidth]{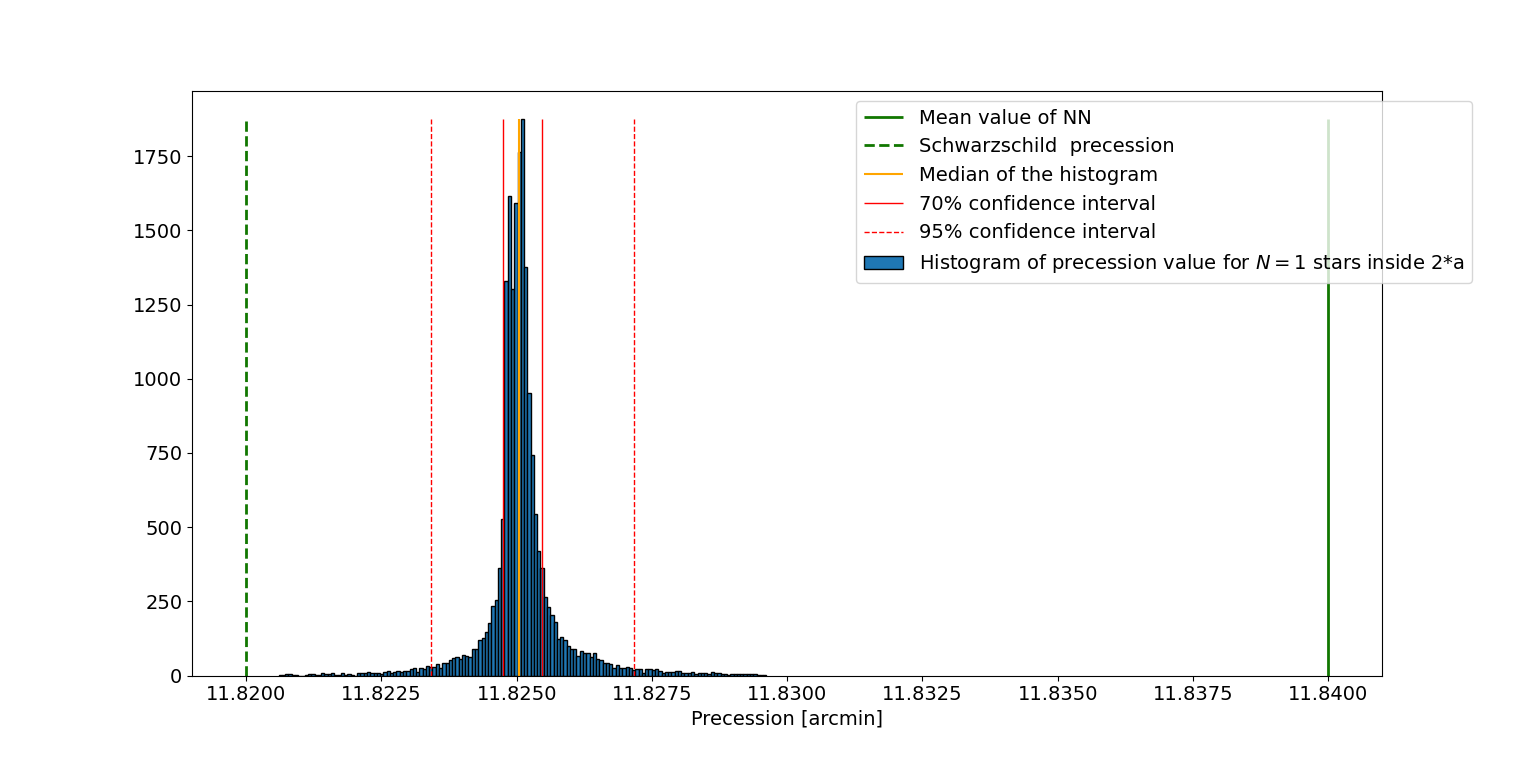}
            \caption{N=1}
        \end{subfigure} &
        \begin{subfigure}[b]{0.45\textwidth}
            \centering
            \includegraphics[width=1\textwidth]{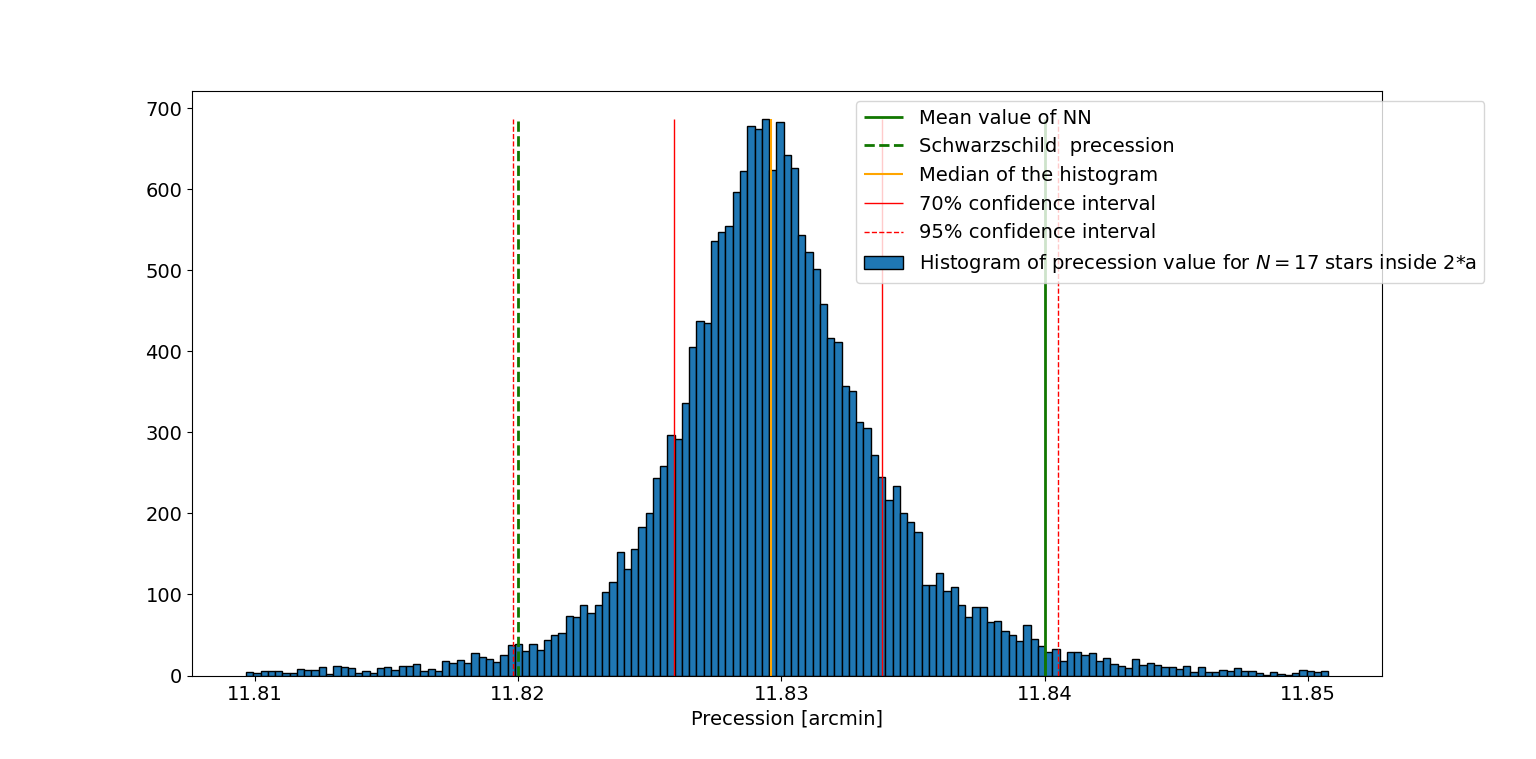}
            \caption{N=17}
        \end{subfigure} \\
        \begin{subfigure}[b]{0.45\textwidth}
            \centering
            \includegraphics[width=1\textwidth]{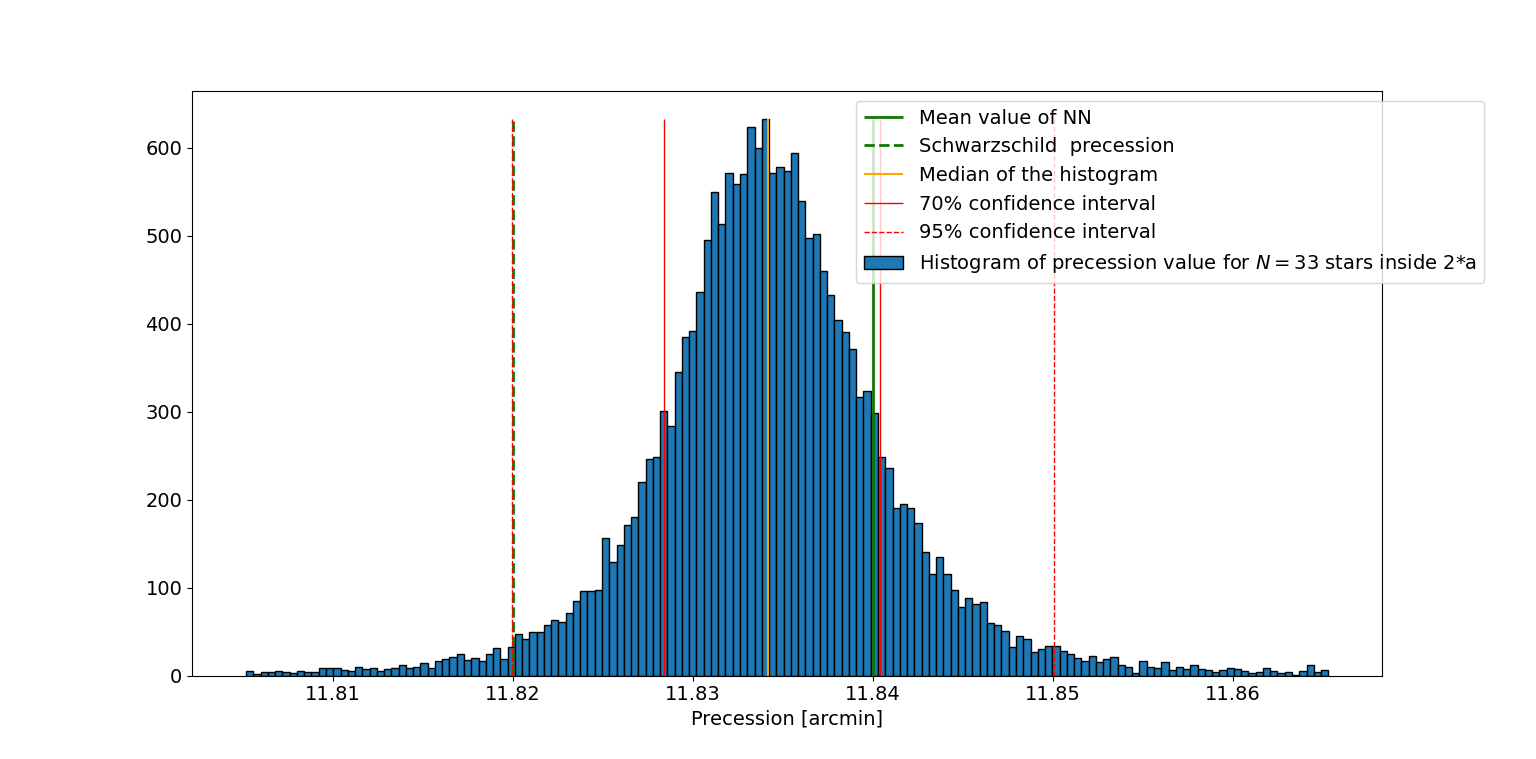}
            \caption{$N=33$}
        \end{subfigure} &
        \begin{subfigure}[b]{0.45\textwidth}
            \centering
            \includegraphics[width=1\textwidth]{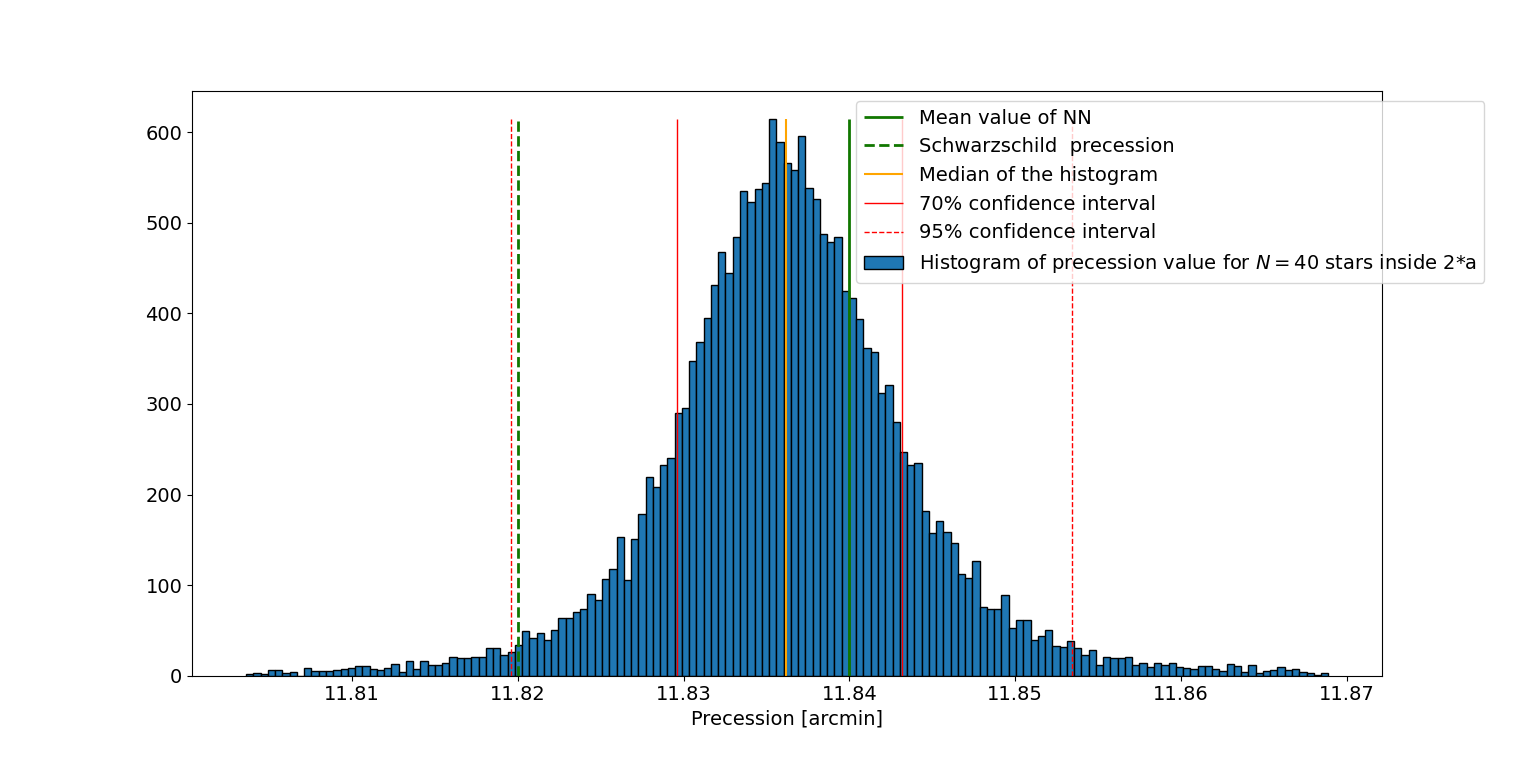}
            \caption{$N = 40$}
        \end{subfigure}
        \\
    \end{tabular}
    \caption{The histograms of $10000$ experiments for $N =  1,\ 17,\ 33,\ 40$. 
    The \textit{green solid} line is the mean value of $\delta\varphi_{\text{Reg}}$, \textit{green dashed } line denotes the Schwarzschild precession calculated via PINN predicted parameters Eq.(\ref{eq:PINN_res}), \textit{orange} line is the median value of the resulting distribution. The \textit{solid red} and \textit{dashed red} lines are the boundaries of the $15\%-85\%$ and $2.5\%-97.5\%$ quantile intervals, respectively.}
    \label{fig:histogram}
\end{figure}

\subsection{Critical concentration}\label{subsec:main}

To find the critical concentration $n_{\text{crit}}$ at which the effect of precession reaches the value $\delta\varphi_{\text{Reg}}$, we consider the following procedure. Within two apocenters $2\hat{a}$, where $\hat{a}=\cfrac{\hat{p}}{1-\hat{e}^2}$, $N$ point masses are generated randomly on which S2 is scattered. For every generated point, the procedure described in Sec. \ref{sec:Scattering} is repeated, and the total precession is taken as the sum of $\delta\varphi_s$ Eq.($\ref{eq:delta_phi}$) for every scattering and the precession of the first-order GR correction in the Schwarzchild metric (Schwarzchild precession) before the scattering $\delta\varphi_{\text{SP}}$. The initial values of focal parameter and eccentricity are taken from PINN 
Eq.(\ref{eq:PINN_res}), then after each scattering are updated according to Sec. \ref{sec:Scattering}. For $N$ from $1$ to $8$ this scheme is repeated $10000$ times and histograms Fig.(\ref{fig:histogram}) are obtained, which show the distribution of the precession value for every $N$. From the histograms it is clear that starting from $N=17$ $\delta\varphi_{\text{Reg}}$ is inside the $95\%$ confidence interval 
and starting from $N_{\text{crit}}=33$ is inside the $70\%$ confidence interval
The confidence intervals are defined as follows: $95\%$ confidence interval when the median of the precession is within the interval  containing $97.5\%$ of the distribution and $2.5\%$ is outside the interval, i.e. is inside between $5\%$ and $95\%$  quantiles; $70\%$ confidence interval when the median of the precession is within the interval of $85\%$ of the distribution and $15\%$ is outside the interval, i.e. is within $15\%$ and $85\%$  quantiles (i.e. the region closest to the peak of the distribution). The value of critical concentration $n_{\text{crit}}$ within 70\% confidence interval is 
\begin{equation}\label{eq:n_crit}
    n_{\text{crit}}\approx N_{\text{crit}}\left(\frac{2^5}{3}\pi \hat{a}^3\right)^{-1}\approx 8.3\times 10^{6}\, pc^{-3}.
\end{equation}

\section{Holtsmark distribution}

According to Eq.(1) in an isotropic system the mean force acting on a star vanishes, \textbf{while the distribution} for a non-vanishing mean square force $\langle\epsilon^2\rangle$ is given as \cite{ChandraStoch,ChNeu,GS}, 
\begin{align}\label{eq:Holtsmark}
    \langle \varepsilon^2 \rangle &= c a^{\frac{4}{3}}\\
    a&=\frac{4}{15}(2\pi G  m)^{\frac{3}{2}}n,
\end{align}
where $m$ is the mean star mass,  $c$ is a dimensionless coefficient 
\begin{equation}\label{eq:c}
    c=\int_0^{\beta_{\text{max}}}\beta^2 H(\beta)d\beta,
\end{equation}
where $\beta$ as the dimensionless per unit mass force is $\varepsilon=\beta a^{\frac{2}{3}}$ \cite{ChandraStoch} and correspondingly $\varepsilon_{\text{max}}=\beta_{\text{max}} a^{\frac{2}{3}}$ is the maximum interaction force during the interaction.

Thus, we assume that at a distance equal to the impact parameter the force acting on S2 is given by the Holtsmark distribution Eq.(\ref{eq:Holtsmark})
\begin{align}
    b_{\text{H}}&:=\left(\frac{G m}{\varepsilon}\right)^{\frac{1}{2}}\\
    \varepsilon &:=\sqrt{\langle\varepsilon^2\rangle}.
\end{align}
For this impact parameter Eq.(\ref{eq:delta_p1}) transforms to the following
\begin{equation}
      \Delta \rho_\text{H} = \frac{-2p Gm}{L\sqrt{1+e^2+2e\cos\varphi_*}} \sqrt{2\pi\left(\frac{4}{15}\right)^{2/3}}c^{1/4}n^{1/3}.
\end{equation}
To find the transformed momentum $\Delta p_{\text{H}}$ we must limit the maximum dimensionless force $\beta_{\text{max}}$ to find the coefficient $c(n)$. As such a force we take the one arising at a distance of $\alpha R_{\odot}$, i.e. (R in [m]): 
\begin{equation*}
    \beta_{\text{max}}(n) \approx 7.8\alpha^{-2}n^{-2/3}\times 10^{-19}. 
\end{equation*}

The evaluation of integral (\ref{eq:c}) yields
\begin{equation*}
    c(n)\approx \frac{10}{\pi} \Gamma\left(\frac{5}{2}\right)\sin\left(\frac{3\pi}{4}\right)\beta^{\frac{1}{2}}_{\text{max}}(n)\approx 3\beta^{\frac{1}{2}}_{\text{max}}(n).
\end{equation*}

Hence,
\begin{equation}\label{eq:rho_H}
    \Delta\rho_\text{H} \approx -0.007 \frac{p GM}{L\sqrt{1+e^2+2e\cos\varphi_*}}\alpha^{-\frac{1}{4}} n^{\frac{1}{4}} .
\end{equation}

\begin{figure}[H]
    \centering
    \includegraphics[width=\linewidth]{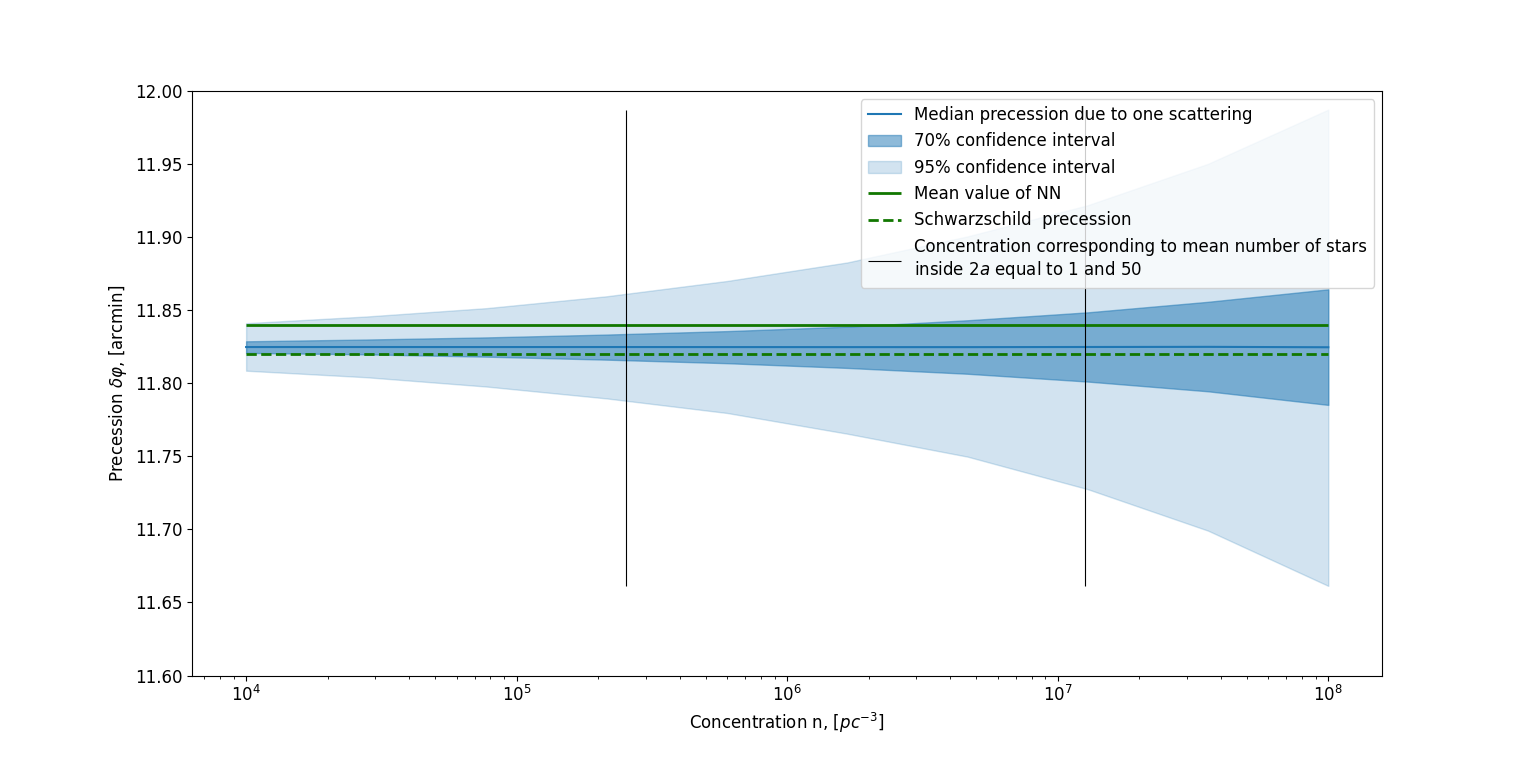}
    \caption{The dependence of the median value (blue solid line) of precession and its 70\% and 95\% confidence intervals from the concentration $n$. The \textit{green solid} line is the mean value of $\delta\varphi_{\text{Reg}}$, \textit{green dashed } line is the Schwarzschild precession calculated via PINN predicted parameters Eq.(\ref{eq:PINN_res}). The solid black lines show the interval of concentrations for which $N=1\div50$}
    \label{fig:holts_n}
\end{figure}

\begin{figure}[H]
    \centering
    \includegraphics[width=\linewidth]{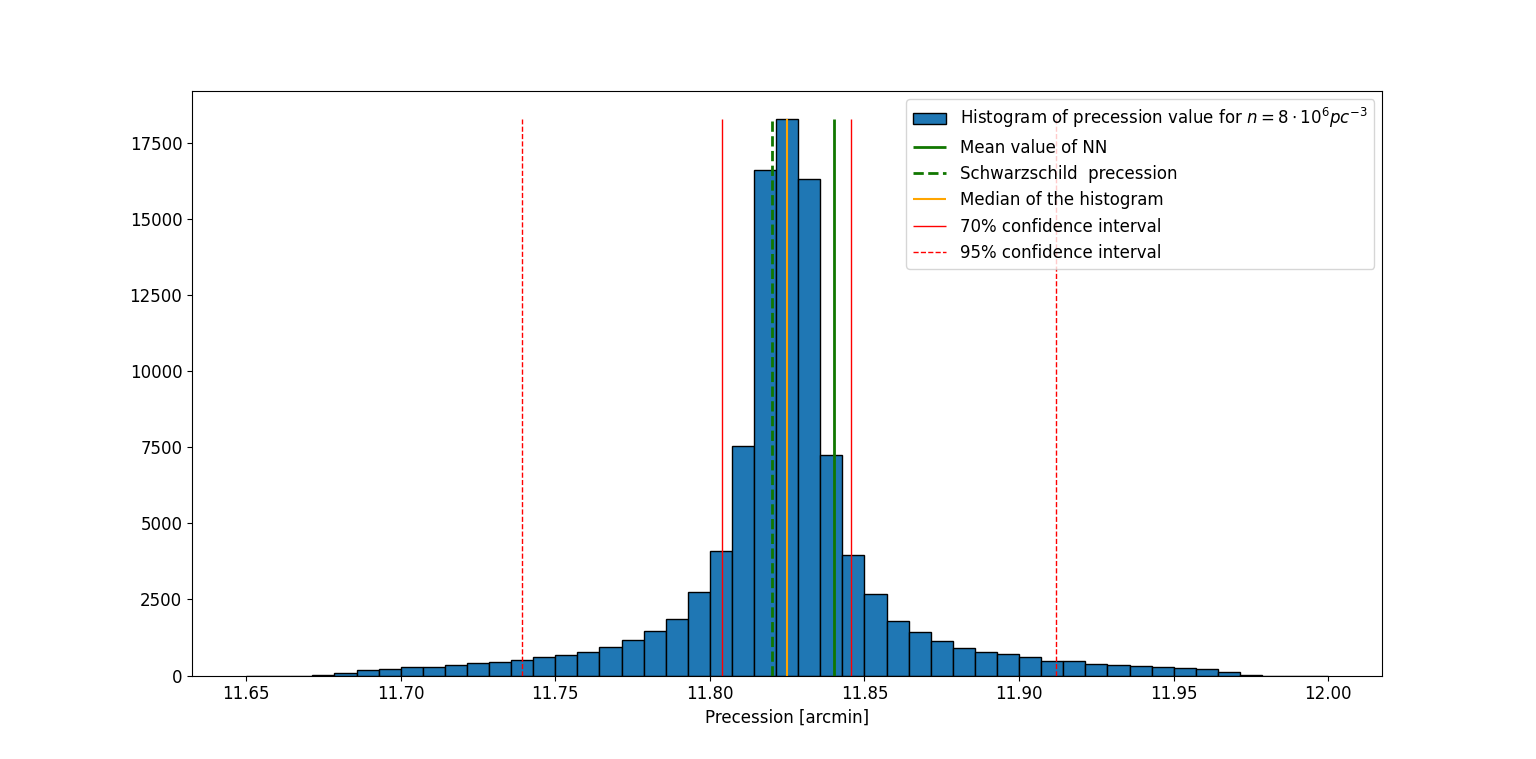}
    \caption{The histograms of precession values obtained by scattering according to the Holtsmark distribution. The \textit{green solid} line is the mean value of $\delta\varphi_{\text{Reg}}$, \textit{green dashed } line is the Schwarzschild precession calculated via PINN predicted parameters Eq.(\ref{eq:PINN_res}), \textit{orange} line is the median value of the resulting distribution. The \textit{solid red} and \textit{dashed red} lines are the boundaries of the $15\%-85\%$ and $2.5\%-97.5\%$ quantile intervals, respectively.}
    \label{fig:holts}
\end{figure}

Calculating the total precession using Eq.(\ref{eq:rho_H}) for concentration values $n=10^4\div10^8\, pc^{-3}$ and assuming that physically significant are those values for $n$ for which the mean number of stars $N$ inside $2\hat{a}$ is $N=1\div50$ we can check whether the obtained values are in agreement with the Monte-Carlo approach in Section \ref{subsec:main}. Fig.(\ref{fig:holts_n}) shows the $70\%$ and $95\%$ bounds of precession value with respect to $n$ and Fig.(\ref{fig:holts}) shows the distribution of precession for $n=8\times 10^6\, pc^{-3}$.

Fig. (\ref{fig:holts_n} and \ref{fig:holts}) show that the procedure described in Section \ref{subsec:main} fully agrees with the disturbance of the orbit according to the Holtsmark distribution.

The massive black hole situated in a star cluster is predicted to lead to a redistribution of the density run of the cluster as $n(r)\propto r^{-7/4}$ \cite{BW} due to the loss of stars at their tidal disruption \cite{GO,R}. The constraint obtained here on the star density near the vicinity of the black hole may support more accurate consideration of the evolutionary effects in the Galactic center core cluster.

\section{Conclusions}

We used the S2-star's orbital data to probe the parameters of the star cluster surrounding the supermassive black hole Sgr A* in the center of the Galaxy. We considered the influence of scatterings on the value of the S2-star precession and the precession value previously found using physics-informed neural networks (PINN). Based on residual precession $\delta\varphi_{\text{S}}$, we found the maximal star density $n_{\text{crit}}$ at which the scattering effect will not exceed the precession previously obtained within PINN analysis $\delta\varphi_{\text{Reg}}$.

We applied the small-angle scattering approximation to find the residual precession $\delta\varphi_{\text{S}}$. Within this approximation and for Monte Carlo random positions inside $2\hat{a}$, the role of S2 and $N$ star scatterings has been analysed. The total precession was calculated as residual to the Schwarzchild precession vs the cumulative effect of the angle shifts at $N$ scatterings. We obtained the critical number of the scatterings, $N_{\text{crit}}=33$, with the mean value of $\delta\varphi_{\text{Reg}}$ inside the $70\%$ confidence interval (between $15\%$ and $85\%$ quantiles), and from there obtained the maximal star density $n_{\text{crit}}\approx 8.3\times 10^{6}\, pc^{-3}$. The results correspond to the stationary distribution of the force per unit mass, i.e  Holtsmark distribution. 
The performed analysis confirmed the efficiency of  PINN in the analysis of the S-star data in further probing of the physical processes in the Galactic center. 

\section{Acknowledgments}
We are thankful to the referee for helpful comments. Sh.K. is acknowledging the ANSEF grant 23AN:PS-astroth-2922.

\section{Data Availability Statement} 
Data sharing not applicable to this article as no datasets were generated or analysed during the current study.

\newpage
\end{document}